\documentclass[a4paper,pre,showpacs,twocolumn]{revtex4}

\usepackage{dcolumn}
\usepackage{graphicx}
\usepackage{ifpdf}
\usepackage{url}

\ifpdf
	\usepackage{pdfsync}
\else
\fi

\usepackage{xspace}

\newcommand{\ie}{i.e.}
 
\newcommand{\eg}{e.g.}
\newcommand{\brim}{BRIM\xspace}

\newcommand{\largeparens}[1]{\left ( #1 \right )}
\newcommand{\largesquare}[1]{\left [ #1 \right ]}
\newcommand{\largecurly}[1]{\left \{ #1 \right \}}
\newcommand{\largeangle}[1]{\left \langle #1 \right \rangle }
\newcommand{\largestraight}[1]{\left | #1 \right |}

\newcommand{\approaches}{\rightarrow}
\newcommand{\by}{\times}

\newcommand{\krondelta}[2]{\ensuremath{\delta\largeparens{#1,#2}}} 
\newcommand{\set}{\largecurly}
\newcommand{\abs}{\largestraight}

\newcommand{\mathpuncspace}{\quad}
\newcommand{\mathcomma}{\mathpuncspace,}
\newcommand{\mathperiod}{\mathpuncspace.}

\newcommand{\mat}[1]{\mathbf{#1}}
\newcommand{\vctr}[1]{\mathbf{#1}}
\newcommand{\innerproduct}[2]{\largeangle{#1, #2}}
\newcommand{\outerproduct}[2]{#1\transpose{#2}} 
\newcommand{\transpose}[1]{#1^\mathrm{T}}
\newcommand{\matrixdims}[2]{\ensuremath{#1 \by #2}}

\newcommand{\trace}[1]{\mathrm{Tr}\,\,#1}
\newcommand{\tracequad}[2]{\traceweightedinner{#1}{#2}{#1}}
\newcommand{\traceweightedinner}[3]{\trace{\weightedinner{#1}{#2}{#3}}}
\newcommand{\weightedinner}[3]{\transpose{#1}#2#3} 

\newcommand{\matrixinnerproduct}[2]{\transpose{#1}{#2}}
\newcommand{\traceinner}[2]{\trace{\matrixinnerproduct{#1}{#2}}}

\newcommand{\zeromat}{\mat{O}}

\newcommand{\matrixbrackets}{\largesquare}
\newcommand{\bipartsubstructurewithsize}[3]{\matrixbrackets{ 
	\begin{array}{cc}
		\zeromat_{\matrixdims{#2}{#2}} & #1_{\matrixdims{#2}{#3}} \\
		(\transpose{#1})_{\matrixdims{#3}{#2}} & \zeromat_{\matrixdims{#3}{#3}}
	\end{array}}}
\newcommand{\bipartsubstructure}[1]{\matrixbrackets{
	\begin{array}{cc}
		\zeromat & #1 \\
		\transpose{#1} & \zeromat
	\end{array}}}
\newcommand{\bipartvctr}[2]{\matrixbrackets{
	\begin{array}{c}
		#1 \\
		#2
	\end{array}}}

\newcommand{\redsum}[1]{\sum_{#1=1}^{\numred}} 
\newcommand{\bluesum}[1]{\sum_{#1=1}^{\numblue}}
\newcommand{\redbluesum}[2]{\redsum{#1}\bluesum{#2}}

\newcommand{\numvertices}{\ensuremath{n}}
\newcommand{\numedges}{\ensuremath{m}}
\newcommand{\numred}{\ensuremath{p}} 
\newcommand{\numblue}{\ensuremath{q}} 

\newcommand{\adjmat}{\mat{A}}
\newcommand{\adjelem}[2]{A_{#1#2}}
\newcommand{\adjsubmat}{\mat{\tilde{A}}} 
\newcommand{\adjsubelem}[2]{\tilde{A}_{#1#2}}

\newcommand{\probmat}{\mat{P}}
\newcommand{\probelem}[2]{P_{#1#2}}
\newcommand{\probsubmat}{\mat{\tilde{P}}} 
\newcommand{\probsubelem}[2]{\tilde{P}_{#1#2}}

\newcommand{\nummodules}{\ensuremath{c}}
\newcommand{\modularity}{\ensuremath{Q}}
\newcommand{\modularitynorm}{\frac{1}{2\numedges}}
\newcommand{\modularitybipartnorm}{\frac{1}{\numedges}}

\newcommand{\module}[1]{\ensuremath{g_{#1}}} 
\newcommand{\redmodule}[1]{\module{#1}}
\newcommand{\bluemodule}[1]{\ensuremath{h_{#1}}}

\newcommand{\modsum}[1]{\sum_{#1=1}^{\nummodules}}

\newcommand{\indmat}{\mat{S}}
\newcommand{\indvec}[1]{\vctr{s}_{#1}} 
\newcommand{\redindmat}{\mat{R}}
\newcommand{\blueindmat}{\mat{T}}
\newcommand{\redindvec}[1]{\vctr{r}_{#1}}
\newcommand{\blueindvec}[1]{\vctr{t}_{#1}}
\newcommand{\redindelem}[2]{R_{#1#2}}
\newcommand{\blueindelem}[2]{T_{#1#2}}

\newcommand{\modmat}{\mat{B}}
\newcommand{\modelem}[2]{B_{#1#2}} 
\newcommand{\modsubmat}{\mat{\tilde{B}}} 
\newcommand{\modsubelem}[2]{\tilde{B}_{#1#2}} 

\newcommand{\reddegrees}[1]{k_{#1}}
\newcommand{\bluedegrees}[1]{d_{#1}}

\newcommand{\normconst}{\ensuremath{C}} 

\newcommand{\leadingevec}{\vctr{x}}
\newcommand{\leadingevecelem}[1]{x_{#1}} 
\newcommand{\leadingeval}{\lambda}
\newcommand{\bipartitionindex}[1]{\largesquare{#1{1} \mid #1{2}}}

\newcommand{\bipartevec}[1]{\vctr{x}_{#1}}
\newcommand{\biparteval}[1]{\lambda_{#1}}
\newcommand{\redevec}[1]{\vctr{u}_{#1}}
\newcommand{\blueevec}[1]{\vctr{v}_{#1}}

\newcommand{\singval}[1]{\sigma_{#1}} 

\newcommand{\redmodwtmat}{\mat{\tilde{R}}}

\newcommand{\redmodwtelem}[2]{\tilde{R}_{#1#2}}
\newcommand{\bluemodwtmat}{\mat{\tilde{T}}}

\newcommand{\bluemodwtelem}[2]{\tilde{T}_{#1#2}}

\newcommand{\nummodulesmax}{\nummodules_\mathrm{max}}

\newcommand{\problinkintra}{p_{\mathrm{in}}}
\newcommand{\problinkinter}{p_{\mathrm{out}}}
\newcommand{\nummodelnetmodules}{N_{\mathrm{mod}}}
\newcommand{\nummodelnetredpermod}{N_{\mathrm{red}}}
\newcommand{\nummodelnetbluepermod}{N_{\mathrm{blue}}}

\newcommand{\mutinfosymbol}{\ensuremath{I}}
\newcommand{\mutinfo}[2]{\mutinfosymbol\largeparens{#1, #2}}
\newcommand{\normalizedmutinfosymbol}{\ensuremath{I_\mathrm{norm}}}
\newcommand{\normalizedmutinfo}[2]{\normalizedmutinfosymbol\largeparens{#1, #2}}
\newcommand{\entropysymbol}{\ensuremath{H}}
\newcommand{\entropy}[1]{\entropysymbol\largeparens{#1}}
\newcommand{\relentsum}[3]{\sum_{#3} #1 \log \frac{#1}{#2}}
\newcommand{\entropysum}[2]{-\sum_{#2} #1 \log #1}
\newcommand{\confusionmat}{\mat{N}}
\newcommand{\confusionelem}[2]{N_{#1 #2}}
\newcommand{\schemex}{\ensuremath{X}}
\newcommand{\schemey}{\ensuremath{Y}}
\newcommand{\indmatx}{\indmat_{X}}
\newcommand{\indmaty}{\indmat_{Y}}
\newcommand{\probability}[1]{\ensuremath{P\largeparens{#1}}}

\newcommand{\wraprefprepost}[3]{\wrapprepost{#1}{#2}{\ref{#3}}}
\newcommand{\formatrefplain}{\ref}
\newcommand{\formatrefparens}{\wraprefprepost{(}{)}}

\newcommand{\wrapprepost}[3]{{#1}{#3}{#2}}

\newcommand{\tagwithlabel}[2]{#1~#2}

\newcommand{\makelabeledcrossrefmacro}[4]
	{\newcommand{#3}{#1{#4}{#2}}}
\newcommand{\makecrossrefmaker}[3]
	{\newcommand{#1}{\makelabeledcrossrefmacro{#2}{#3}}}

\newcommand{\eqnrefformat}{\formatrefparens}
\newcommand{\eqnlabelbinding}{\tagwithlabel}
\newcommand{\eqnlabel}{eq.}
\newcommand{\Eqnlabel}{Eq.}
\newcommand{\eqnslabel}{eqs.}
\newcommand{\Eqnslabel}{Eqs.}

\newcommand{\eqnnum}{\eqnrefformat}
\makecrossrefmaker{\newlabeledeqnref}{\eqnlabelbinding}{\eqnnum}
\makecrossrefmaker{\newwordpluseqnref}{\tagwithlabel}{\eqnnum}

\newlabeledeqnref{\eqn}{\eqnlabel}
\newlabeledeqnref{\Eqn}{\Eqnlabel}
\newlabeledeqnref{\eqns}{\eqnslabel}
\newlabeledeqnref{\Eqns}{\Eqnslabel}

\newwordpluseqnref{\andeqn}{and}
\newwordpluseqnref{\througheqn}{through}

\newcommand{\figrefformat}{\formatrefplain}
\newcommand{\figlabelbinding}{\tagwithlabel}
\newcommand{\figlabel}{fig.}
\newcommand{\Figlabel}{Fig.}
\newcommand{\figslabel}{figs.}
\newcommand{\Figslabel}{Figs.}

\newcommand{\fignum}{\figrefformat}
\makecrossrefmaker{\newlabeledfigref}{\figlabelbinding}{\fignum}
\makecrossrefmaker{\newwordplusfigref}{\tagwithlabel}{\fignum}

\newlabeledfigref{\fig}{\figlabel}
\newlabeledfigref{\Fig}{\Figlabel}
\newlabeledfigref{\figs}{\figslabel}
\newlabeledfigref{\Figs}{\Figslabel}

\newwordplusfigref{\andfig}{and}
\newwordplusfigref{\throughfig}{through}

\newcommand{\sxnrefformat}{\formatrefplain}
\newcommand{\sxnlabelbinding}{\tagwithlabel}
\newcommand{\sxnlabel}{section}
\newcommand{\Sxnlabel}{Section}
\newcommand{\sxnslabel}{sections}
\newcommand{\Sxnslabel}{Sections}

\newcommand{\sxnnum}{\sxnrefformat}
\makecrossrefmaker{\newlabeledsxnref}{\sxnlabelbinding}{\sxnnum}
\makecrossrefmaker{\newwordplussxnref}{\tagwithlabel}{\sxnnum}

\newlabeledsxnref{\sxn}{\sxnlabel}
\newlabeledsxnref{\Sxn}{\Sxnlabel}
\newlabeledsxnref{\sxns}{\sxnslabel}
\newlabeledsxnref{\Sxns}{\Sxnslabel}
	
\newwordplussxnref{\andsxn}{and}
\newwordplussxnref{\throughsxn}{through}

\newcommand{\tblrefformat}{\formatrefplain}
\newcommand{\tbllabelbinding}{\tagwithlabel}
\newcommand{\tbllabel}{table}
\newcommand{\Tbllabel}{Table}
\newcommand{\tblslabel}{tables}
\newcommand{\Tblslabel}{Tables}

\newcommand{\tblnum}{\tblrefformat}
\makecrossrefmaker{\newlabeledtblref}{\tbllabelbinding}{\tblnum}
\makecrossrefmaker{\newwordplustblref}{\tagwithlabel}{\tblnum}

\newlabeledtblref{\tbl}{\tbllabel}
\newlabeledtblref{\Tbl}{\Tbllabel}
\newlabeledtblref{\tbls}{\tblslabel}
\newlabeledtblref{\Tbls}{\Tblslabel}
	
\newwordplustblref{\andtbl}{and}
\newwordplustblref{\throughtbl}{through}
\renewcommand{\eqnlabel}{Eq.}
\renewcommand{\eqnslabel}{Eqs.}
\renewcommand{\Eqnlabel}{Equation}
\renewcommand{\Eqnslabel}{Equations}
\renewcommand{\figlabel}{Fig.}
\renewcommand{\figslabel}{Figs.}
\renewcommand{\Figlabel}{Figure}
\renewcommand{\Figslabel}{Figures}

\graphicspath{{pdffigs/}{epsfigs/}{mpsfigs/}}

\begin{document}

% FIXME: boring title
\title{Modularity and community detection in bipartite networks.  }

\author{Michael J. Barber}

\affiliation{Austrian Research Centers GmbH---ARC, Bereich systems research, Vienna, Austria}
\email{michael.barber@arcs.ac.at}

\date{\today}

\begin{abstract}
	
The modularity of a network quantifies the extent, relative to a null model network, to which vertices cluster into community groups. We define a null model appropriate for bipartite networks, and use it to define a bipartite modularity. The bipartite modularity is presented in terms of a modularity matrix \( \modmat \); some key properties of the eigenspectrum of \( \modmat \) are identified and used to describe an algorithm for identifying modules in bipartite networks. The algorithm is based on the idea that the modules in the two parts of the network are dependent, with each part mutually being used to induce the vertices for the other part into the modules. We apply the algorithm to real-world network data, showing that the algorithm successfully identifies the modular structure of bipartite networks.

\end{abstract}


\maketitle

\section{Introduction}\label{sec:introduction} % (fold)

Networks have attracted a burst of attention in the last decade (useful reviews include Refs.~\cite{ChrAlb:2007,New:2006,New:2003,AlbBar:2002}), with applications to natural, social, and technological networks. Of great current interest is the identification of the modular structure of the network. Detecting modules, or communities, allows quantitative investigation of relevant subnetworks, which may have different properties from the aggregate properties of the network as a whole, \eg,  modules in the World Wide Web are sets of topically related web pages.

Informally, a network module is a subgraph whose vertices are more likely to be connected to one another than to the vertices outside the subgraph. A variety of approaches \citep{AngBocMarPelStr:2007,GolKog:2006,Has:2006,NewLei:2007,ReiBor:2006,PalDerFarVic:2005,NewGir:2004,ClaNewMoo:2004,GirNew:2002} have been taken to explore this concept. See Refs.~\cite{DanDiaDucAre:2005,New:2004b} for useful reviews. 

In this work we focus on the measure called modularity, introduced by \citet{NewGir:2004}. Modularity reflects the extent, relative to a null model network, to which edges are formed within modules instead of between modules. Using the modularity, we can assess the quality of any assignment of vertices to modules. Further, the module identification problem becomes a modularity optimization problem. However, exact maximization of the modularity is in general an intractable problem, because the number of ways to partition the set of vertices grows extremely rapidly \citep{Rot:1964}. In light of this, a number of effective algorithms have been introduced to find high modularity partitions of the vertices \citep{PujBejDel:2006,New:2004a}. The modularity can be also be defined in terms of a so-called modularity matrix, the eigenspectrum of which has a fundamental relationship with the modular nature of the network \citep{New:2006a}.

Given the explicit dependence of the modularity upon a null model, it is clear that the specific choice of null model has a profound impact on the modularity. Surprisingly, only one null model has been so far  explored at length: networks with edges randomly assigned such that the expected degrees of model-network vertices equal the actual degrees of corresponding real-network vertices \citep{New:2006a}. Specific classes of networks have additional constraints that could be and, indeed, should be reflected in the null model.

A significant such class of networks is that of bipartite networks. The vertices of a bipartite network can be partitioned into two disjoint sets such that no two vertices within the same set are adjacent. There are thus two distinct kinds of vertices, providing a natural representation for many affiliation or interaction networks, with one kind of vertex representing actors and the other representing relations. Examples of actor-relation pairs include people attending events \citep{DavGarGar:1941,Fre:2003,DorBatFer:2004}, court justices making decisions \citep{DorBatFer:2004}, scientists jointly publishing articles \citep{New:2001a,New:2001}, organizations collaborating in projects \citep{BarKruKruRoe:2006,RoeBar:2007}, and legislators serving on committees \citep{PorMucNewFri:2007}. Arguably, bipartite networks are the empirically standard case for social networks and other interaction networks, with unipartite networks appearing---often implicitly---as projections.

In the statistical physics community, the usual approach taken to identify modules in bipartite networks is to first construct a unipartite projection of one part of the network, and then identify modules in that projection using methods for unipartite networks. For example, in the scientist-publication network mentioned above, a network of scientists is created by linking scientists when they have jointly published. These unipartite projection can be illuminating, but intrinsically lose information---indeed, \citet{GuiSalAma:2007} demonstrate that analysis of an unweighted, unipartite projection can give unreliable or incorrect results.

The principal contribution in this work is a proposed definition of a modularity for bipartite networks. The approach taken is based on defining a bipartite modularity matrix \( \modmat \) as an extension of the recent work by \citet{New:2006a}. Some key properties of the eigenspectrum of \( \modmat \) are identified and used to specialize Newman's matrix-based algorithms to bipartite networks. An additional algorithm fundamentally based on the bipartite character of the networks is introduced; we call the algorithm \brim, for bipartite, recursively induced modules.

In parallel, \citet{GuiSalAma:2007} have independently investigated modularity in bipartite networks. They proceed by first identifying the two parts of the network as actors and teams, and then formulating a bipartite modularity in which modules consist of groups of actors that are closely interconnected based on joint participation in many teams. The resulting modularity is thus focused on identifying modules in only one part of the network at a time. Interesting, \citeauthor{GuiSalAma:2007} point out the possibilities of classifying both partite sets of the network simultaneously and of customizing spectral methods for bipartite networks, which is essentially the approach taken in the present work.  

As of this writing, we are aware of no other attempts to define modularity for bipartite networks. However bipartite networks, or ``two mode networks,'' have undergone several related studies in the sociology community using other methods (see, \eg, Refs.~\cite{DorBatFer:2004,Fre:2003} and references cited therein).

The structure of the paper is as follows: in \sxn{sec:bipartite_modularity} we define a modularity matrix and measure for bipartite networks. We discuss using the bipartite modularity matrix to identify modules in \sxn{sec:module_identification}, and apply the algorithm therein devised to two real-world networks in \sxn{sec:results}. Finally, we conclude in \sxn{sec:conclusions} with an assessment of the present investigation and an outlook for future work.

% section introduction (end)

\section{Bipartite Modularity}\label{sec:bipartite_modularity} % (fold)

In this section, we develop a modularity matrix for bipartite networks. Structurally and notationally, the development parallels the discussion of the modularity matrix by \citet{New:2006a}.

Consider a network with \( \numvertices \) vertices and \( \numedges \) edges defined by an adjacency matrix \( \adjmat \). Each vertex \( i \) is assigned to a community group or module, denoted by \module{i}. The modularity \( \modularity \) for such an assignment reflects the extent, relative to a null model, to which edges are formed within modules instead of between modules. Formally, the modularity is defined as
\begin{equation}
	\modularity = \modularitynorm \sum_{i,j} 
			\largeparens{\adjelem{i}{j} - \probelem{i}{j}} 
			\krondelta{\module{i}}{\module{j}}
	\mathcomma
	\label{eq:defmodularity}
\end{equation}
where the \( \adjelem{i}{j} \) are the adjacency matrix elements and the \( \probelem{i}{j} \) are probabilities in the null model that an edge exists between vertices \( i \) and \( j \).

The modularity can be given an equivalent definition in matrix form. First, the \( \numvertices \) community indices \( \module{i} \) with values taken from \( \set{1, 2, \ldots, \nummodules} \) are replaced by an \( \matrixdims{\numvertices}{\nummodules} \) index matrix \( \indmat = \largesquare{\indvec{1} \mid \indvec{2} \mid \cdots \mid \indvec{\nummodules}} \), where \( \nummodules \) is the number of modules. All elements of \( \indmat \) take on either a 0 or 1 value, so that column \( \indvec{i} \) is an index vector showing membership in module \( i \); a value of 1 in position \( j \) of \( \indvec{i} \) indicates that vertex \( j \) belongs to module \( i \). Given that each vertex is assigned to exactly one module, each row of \( \indmat \) has a single unit value and the index vectors are thus orthogonal. 

Further, a modularity matrix \( \modmat \) is defined with elements 
\begin{equation}
	\modelem{i}{j} = \adjelem{i}{j} - \probelem{i}{j}
	\mathperiod
	\label{eq:defmodularitymatrix}
\end{equation}
Using \( \indmat \) and \( \modmat \), the modularity becomes
\begin{equation}
	\modularity = \modularitynorm\tracequad{\indmat}{\modmat}
	\mathperiod
	\label{eq:defmatrixformmodularity}
\end{equation}
The eigenspectrum of \( \modmat \) has a fundamental relationship with the modular nature of the network, as \citet{New:2006a} has explored. 

From \eqns{eq:defmodularity} \througheqn{eq:defmatrixformmodularity}, it is apparent that the choice of null model has a profound impact on the modularity. Thus, for example, a Bernoulli random graph with constant \( \probelem{i}{j} = p\) for all \( i \) and \( j \) is a poor representation of most real-world networks, so would be an inappropriate choice of null model. Instead, the usual choice of null model \citep{New:2006a} assigns edges at random with the expected degrees of model vertices constrained to match the degrees in the actual network. 

In much the same fashion, bipartite networks have specific constraints that should be reflected in the null model. The vertices of a bipartite network can be partitioned into two disjoint sets such that no two vertices within the same set are adjacent. An equivalent, but more visual, definition is that the vertices in a bipartite graph can be assigned one of two colors, say red and blue, with no neighboring vertices bearing the same color. In the remainder of this section, we will define a null model with the above requirement that the expected degrees match the degrees in the real network, along with the additional constraint that each edge links a red vertex and a blue vertex.

Let \( \numred \) be the number of red vertices and \( \numblue \) be the number of blue vertices; this implies \( \numvertices = \numred + \numblue \). Without loss of generality, assume that the vertices are indexed so that red vertices are labeled \( 1, 2, \ldots, \numred \) and the blue vertices are labeled \( \numred+1, \numred+2, \ldots, \numred+\numblue \). The adjacency matrix then has a block off-diagonal form of
\begin{equation}
	\adjmat = \bipartsubstructurewithsize{\adjsubmat}{\numred}{\numblue}
	\mathcomma
	\label{eq:adjmatrixsubstructure}
\end{equation}
where \( \zeromat_{\matrixdims{i}{j}} \) is the all-zero matrix with \( i \) rows and \( j \) columns. Require the same block structure for \( \probmat \) that is exhibited by \( \adjmat \), giving
\begin{equation}
	\probmat = \bipartsubstructurewithsize{\probsubmat}{\numred}{\numblue}
	\mathperiod
	\label{eq:probmatrixsubstructure}
\end{equation}
This form for \( \probmat \) assigns zero likelihood to edges between vertices with the same color,  precluding any such edges in the null model.

The modularity matrix \( \modmat \) in turn has a block off-diagonal form of
\begin{equation}
	\modmat = \bipartsubstructurewithsize{\modsubmat}{\numred}{\numblue}
	\mathcomma
	\label{eq:modmatrixsubstructure}
\end{equation}
where \( \modsubmat = \adjsubmat - \probsubmat \). The all-zero blocks on the diagonal are the potential modularity contributions from pairs of vertices of the same color being present in a module; all meaningful contributions, positive or negative, to the modularity thus are made by pairs of vertices with distinct colors. In contrast, with the usual null model based on unipartite networks \citep{New:2006a}, the corresponding blocks contain only negative elements (or zeros for isolated nodes of degree zero), always providing a modularity penalty for pairs of like-colored vertices in the same module.

\Eqn{eq:defmodularity} can be rewritten as
\begin{equation}
	\modularity = \modularitybipartnorm \redbluesum{i}{j} \modsubelem{i}{j} \krondelta{\redmodule{i}}{\bluemodule{j}}
	\label{eq:defsubstructuremodularity}
	\mathcomma
\end{equation}
where \( \bluemodule{j} = \module{j+\numred} \). Since \( \modularity = 0 \) when all vertices are in the same module, we can set all \( \redmodule{i} \) and \( \bluemodule{j} \) equal, giving
\begin{equation}
	\redbluesum{i}{j}\largeparens{\adjsubelem{i}{j} - \probsubelem{i}{j}} = 0
\end{equation}
so that 
\begin{equation}
	\redbluesum{i}{j} \probsubelem{i}{j} = \redbluesum{i}{j} \adjsubelem{i}{j} = \numedges
	\mathperiod
	\label{eq:submatrixsumstonumedges}
\end{equation}
Thus, the expected number of edges in the null model must equal the number of edges in the actual network. 

The degrees of the red vertices are given by \( \bluesum{j}\adjsubelem{i}{j} = \reddegrees{i} \), while those of the blue vertices are given by \( \redsum{i}\adjsubelem{i}{j} = \bluedegrees{j} \). By constraining the expected degrees in the null model to match the actual degrees, as discussed above, we obtain
\begin{eqnarray}
	\bluesum{j}\probsubelem{i}{j} & = & \reddegrees{i} \label{eq:expectedreddegrees}\\
	\redsum{i}\probsubelem{i}{j} & = & \bluedegrees{j} \label{eq:expectedbluedegrees}
	\mathperiod
\end{eqnarray}
Since
\begin{eqnarray}
	\redsum{i} \reddegrees{i} & = & \numedges \label{eq:reddegreesum}\\
	\bluesum{j} \bluedegrees{j} & = & \numedges \label{eq:bluedegreesum}
	\mathcomma
\end{eqnarray}
\eqns{eq:expectedreddegrees} \andeqn{eq:expectedbluedegrees} ensure that \eqn{eq:submatrixsumstonumedges} holds.

In the usual null model, the probability of an edge being present between two vertices is proportional to the product of the degrees of the vertices. For the bipartite case, this becomes \( \probsubelem{i}{j} = \normconst \reddegrees{i} \bluedegrees{j} \) for some constant \( \normconst \). Combining this definition with \eqns{eq:expectedbluedegrees} \andeqn{eq:reddegreesum}, we obtain
\begin{eqnarray}
	\bluedegrees{j} & = & \redsum{i} \probsubelem{i}{j} \nonumber \\
	& = & \normconst \redsum{i} \reddegrees{i} \bluedegrees{j} \nonumber \\
	& = & \largeparens{\normconst \numedges} \bluedegrees{j} \label{eq:derivenormconst}
	\mathcomma
\end{eqnarray}
so that \( \normconst = 1/\numedges \) and thus
\begin{equation}
	\probsubelem{i}{j} = \frac{\reddegrees{i} \bluedegrees{j}}{\numedges}
	\mathperiod
	\label{eq:defprobedges}
\end{equation}
The same result can be obtained from \eqns{eq:expectedreddegrees} \andeqn{eq:bluedegreesum} instead of \eqns{eq:expectedbluedegrees} \andeqn{eq:reddegreesum}. With \eqn{eq:defprobedges}, we have fully defined the modularity \( \modularity \) for a bipartite network.

% section bipartite_modularity (end)

\section{Module Identification}\label{sec:module_identification} % (fold)

\subsection{Spectral Methods for Module Identification}\label{sec:spectral_methods_for_module_identification} % (fold)

Using the modularity defined in \sxn{sec:bipartite_modularity}, we can assess the quality of any partitioning of the vertices of a bipartite graph into modules. A partitioning can be determined using any method. Two general approaches seem relevant. First, the modularity defined in \sxn{sec:bipartite_modularity} can be maximized using standard optimization algorithms such as genetic algorithms, greedy search methods \citep{New:2004a}, or extremal optimization \citep{DucAre:2005}; this is generally straightforward and will not be discussed at length in this work. Second, the spectral properties of \( \modmat \) or other matrices associated with the graph can be analyzed to partition the vertices into modules.

For example, one standard partitioning approach is to assign the vertices to modules using spectral partitioning (SP). In spectral partitioning, the eigenvectors of the network Laplacian are used to minimize the number of edges running between groups. The SP approach has a significant drawback: the vertices are assigned to modules of predetermined size. This is problematic for the investigation of real-world networks, where the number and sizes of community groups are not generally known in advance. 

An analogous approach based on the spectral properties of the modularity matrix \( \modmat \) has recently been proposed \citep{New:2006a}. Since the modularity is conceptually closer to our understanding of network community structure, this spectral optimization of modularity (SOM) is better tailored for real world networks. 

An important special case in both spectral partitioning and spectral optimization of modularity is to assign the vertices to two groups based on a single eigenvector of the Laplacian (SP) or modularity (SOM) matrix. In the case of SP, we are interested in the eigenvector corresponding to the smallest positive eigenvalue; this is the Fiedler vector. For SOM, we are interested in the leading eigenvector \( \leadingevec \), corresponding to the largest positive eigenvalue \( \leadingeval \) of \( \modmat \); we propose calling this the \emph{Newman vector}. Using the Newman vector, we approximate \( \modmat \) as
\begin{equation}
	\modmat \approx \leadingeval \outerproduct{\leadingevec}{\leadingevec}
	\mathperiod
	\label{eq:modularitymatrix1d}
\end{equation} 
With just two modules, \( \indmat = \bipartitionindex{\indvec} \), so that the modularity in \eqn{eq:defmatrixformmodularity} becomes
\begin{equation}
	\modularity = \frac{\leadingeval}{2\numedges} \largeparens{
		\innerproduct{\indvec{1}}{\leadingevec}^{2} + 
		\innerproduct{\indvec{2}}{\leadingevec}^{2}}
	\mathperiod
	\label{eq:modularity1d}
\end{equation}
Recall that the index vectors \( \indvec{1} \) and \( \indvec{2} \) take on values from \( \set{0, 1} \). It is clear how to maximize the modularity in \eqn{eq:modularity1d}: when \( \leadingevecelem{i} \), the \(i\)th element of \(\leadingevec\), is positive, assign vertex \( i \) to the first module by setting the \( i \)th entry of \( \indvec{1} \) to one, and when \( \leadingevecelem{i} \) is negative, assign vertex \( i \) to the second module by setting the \( i \)th entry of \( \indvec{2} \) to one \footnote{The assignment when \( \leadingevecelem{i} = 0 \) is arbitrary, and makes no contribution to the modularity.}.

The use of multiple of eigenvectors allows more than two modules to be considered (\( \nummodules > 2 \)), with at most one module more than the number of positive eigenvalues of \( \modmat \) \citep{New:2006a}. Additional eigenvectors of \( \modmat \) can also be used for SOM \citep{New:2006a} in a vector partitioning algorithm adapted from spectral partitioning \citep{AlpYao:1995,AlpKahYao:1999}. In the present work, we will not make use of this algorithm, nor of a recursive bipartitioning approach, instead developing an alternative technique that capitalizes on the bipartite nature of the networks.

% subsection spectral_methods_for_module_identification (end)

\subsection{Module Identification in Bipartite Networks}\label{sec:module_identification_in_bipartite_networks} % (fold)

In \sxn{sec:spectral_methods_for_module_identification}, we have seen how to identify community groups of networks by using the Newman vector to maximize \( \modularity \). However, we made no use of the bipartite character of the networks. For a bipartite network, the eigenvalue equation \( \modmat \bipartevec{i} = \biparteval{i} \bipartevec{i}\) can be written as 
\begin{equation}
	\bipartsubstructure{\modsubmat} \bipartvctr{\redevec{i}}{\blueevec{i}} =
	\biparteval{i} \bipartvctr{\redevec{i}}{\blueevec{i}}
	\mathcomma
	\label{eq:bipartmodmateigenvalueproblem}
\end{equation}
where \( \redevec{i} \) is a \( \matrixdims{\numred}{1} \) vector and \( \blueevec{i} \) is a \( \matrixdims{\numblue}{1} \) vector. The left-hand side of \eqn{eq:bipartmodmateigenvalueproblem} can be multiplied out, giving
\begin{equation}
	\bipartsubstructure{\modsubmat} \bipartvctr{\redevec{i}}{\blueevec{i}} =
	\bipartvctr{\modsubmat\blueevec{i}}{\transpose{\modsubmat}\redevec{i}} =
	\biparteval{i} \bipartvctr{\redevec{i}}{\blueevec{i}}
	\mathcomma
	\label{eq:bipartmodmateigenvalueproblemexpanded}
\end{equation}
\ie, \( \modsubmat\blueevec{i} = \biparteval{i}\redevec{i}   \) and \( \transpose{\modsubmat}\redevec{i} = \biparteval{i}\blueevec{i} \).

Additionally, we can construct a vector from \( \redevec{i} \)  and \( - \blueevec{i} \), so that
\begin{equation}
	\bipartsubstructure{\modsubmat} \bipartvctr{\redevec{i}}{-\blueevec{i}} =
	\bipartvctr{-\modsubmat\blueevec{i}}{\transpose{\modsubmat}\redevec{i}} =
	-\biparteval{i} \bipartvctr{\redevec{i}}{-\blueevec{i}}
	\mathperiod
	\label{eq:bipartmodmatnegativeeigenvalues}
\end{equation}
Hence, for any eigenvalue \( \biparteval{i} \) of \( \modmat \),  \( - \biparteval{i} \) is also an eigenvalue of \( \modmat \). 

Since only the eigenvectors corresponding to positive eigenvalues of \( \modmat \)  can give positive contributions to \( \modularity \), we can focus on just the positive eigenvalues \( \singval{i} = \abs{\biparteval{i}} > 0\). In this case, \( \redevec{i} \) and \( \blueevec{i} \) are, respectively, left and right singular vectors of \( \modsubmat \). If we shift our attention from the spectral decomposition of \( \modmat \) to the singular value decomposition (SVD) of \( \modsubmat \), we therefore automatically exclude the eigenvectors of \( \modmat \) that correspond to negative eigenvalues.

The appearance of the singular vectors of \( \modsubmat \) is not surprising. All the information about the linkage structure of the network is contained in \( \modsubmat \), and the singular value decomposition is the natural generalization of the spectral decomposition used for \( \modmat \) to asymmetric matrices like \( \modsubmat \). What is more, the singular values and singular vectors of \( \modsubmat \) can sometimes provide more information than the eigenvalues and eigenvectors of \( \modmat \). 

For example, the number of modules is at most one more than the number of positive eigenvalues of \( \modmat \). Since, for each vertex, the expected degree in the null model equals the actual degree in the network, the rows and columns of \( \modsubmat \) all sum to zero. The rank \( r \) of \( \modsubmat \), which equals the number of singular values of \( \modsubmat \), must then be less than both \( \numred \) and \( \numblue \). From this, we conclude that the number of communities is at most equal to the smaller of \( \numred \) and \( \numblue \). 

To assign vertices to modules using \( \modsubmat \), we first partition the index matrix \( \indmat \) so that
\begin{equation}
	\indmat = \bipartvctr{\redindmat}{\blueindmat}
	\mathperiod
	\label{eq:partitionindexmatrix}
\end{equation}
The matrices \( \redindmat \) and \( \blueindmat \) have dimensions \( \matrixdims{\numred}{\nummodules} \) and \( \matrixdims{\numblue}{\nummodules} \), respectively, indexing the red and blue vertices into \( \nummodules \) modules. 
% The columns of \( \redindmat \) and \( \blueindmat \) are partitioned index vectors \( \redindvec{i} \) and \( \blueindvec{i} \) satisfying
% \begin{equation}
% 	\indvec{i} = \bipartvctr{\redindvec{i}}{\blueindvec{i}}
% 	\mathperiod
% 	\label{eq:partitionindexvector}
% \end{equation}
Substituting the partitioned matrices into \eqn{eq:defmatrixformmodularity}, we obtain
\begin{equation}
	\modularity = \modularitybipartnorm \traceweightedinner{\redindmat}{\modsubmat}{\blueindmat}
	\mathperiod
	\label{eq:defsubmatmodularity}
\end{equation}
Our goal then becomes to assign network vertices to modules such that \eqn{eq:defsubmatmodularity} is maximized. 

One approach to optimizing the modularity as expressed in \eqn{eq:defsubmatmodularity} is essentially the same as the Newman vector approach considered in \sxn{sec:spectral_methods_for_module_identification}. Without loss of generality, label the singular values such that \( \singval{1} \geq \singval{2} \geq \singval{r} > 0 \). Approximate \( \modsubmat \) as
\begin{equation}
	\modsubmat \approx \singval{1}\outerproduct{\redevec{1}}{\blueevec{1}}
	\mathperiod
	\label{eq:leadingsingvalmodularityapproximation}
\end{equation}
Now, we bipartition the vertices with \( \redindmat = \bipartitionindex{\redindvec} \) and \( \blueindmat = \bipartitionindex{\blueindvec} \), so that
\begin{equation}
	\modularity = \frac{\singval{1}}{\numedges} \largeparens{
		\innerproduct{\redindvec{1}}{\redevec{1}}
		\innerproduct{\blueindvec{1}}{\blueevec{1}} 
		+
		\innerproduct{\redindvec{2}}{\redevec{1}}
		\innerproduct{\blueindvec{2}}{\blueevec{1}} 
	}
	\mathperiod
	\label{eq:bipartmodularity1d}
\end{equation}
As with the Newman vector approach, \( \modularity \) is maximized by assigning the vertices to modules based on the signs of the corresponding component of \( \redevec{1} \) or \( \blueevec{1} \), as appropriate. This maximizes the magnitude of the inner products in \eqn{eq:bipartmodularity1d}, with consistent assignment of both red and blue vertices to the same module based on the signs ensuring that positive contributions are made to the modularity.

% subsection module_identification_in_bipartite_networks (end)

\subsection{Recursive Identification of Bipartite Modules}\label{sec:recursive_identification_of_bipartite_modules} % (fold)

In \sxns{sec:spectral_methods_for_module_identification} \andsxn{sec:module_identification_in_bipartite_networks}, we have seen how the leading eigenvector of \( \modmat \) and the leading singular vectors of \( \modsubmat \) can be used to bipartition network vertices. Extending these methods to use the full modularity matrices and to handle more than two modules is in general nontrivial. However, for the bipartite case at least, there is a relatively straightforward extension that leads to a useful algorithm.

First, we assume that the blue vertices are all assigned to modules through some mechanism. Maximizing the modularity then consists solely of assigning the red vertices to modules. This is a comparatively simple task. To see this, rewrite \eqn{eq:defsubmatmodularity}, giving 
\begin{eqnarray}
	\modularity & = & \modularitybipartnorm \traceweightedinner{\redindmat}{\modsubmat}{\blueindmat} \nonumber\\
	& = & \modularitybipartnorm \traceinner{\redindmat}{\bluemodwtmat} 
	\mathcomma
	\label{eq:fixedbluemodularity}
\end{eqnarray}
where we have aggregated the fixed terms into the matrix \( \bluemodwtmat = \modsubmat \blueindmat \). We now write \eqn{eq:fixedbluemodularity} in terms of explicit sums, so that
\begin{eqnarray}
	\modularity & = & \modularitybipartnorm \modsum{k} \redsum{i} \redindelem{i}{k} \bluemodwtelem{i}{k} \nonumber \\
	& = & \modularitybipartnorm \redsum{i} \largeparens{\modsum{k} \redindelem{i}{k} \bluemodwtelem{i}{k}}
	\mathperiod
	\label{eq:fixedbluemodularityrowsums}
\end{eqnarray}
The inner sum in \eqn{eq:fixedbluemodularityrowsums} is a sum across the rows of \( \redindmat \). Since each row of \( \redindmat \) consists of a single 1 with all other elements being 0, the modularity is now simple to maximize: we just assign red vertex \( i \) to module \( k \) such that \( \bluemodwtelem{i}{k} \) is the maximum of the \( i \)th row of \( \bluemodwtmat \) \footnote{An arbitrary rule is needed to break ties, for example, random assignment of the vertex to one of the modules that maximizes \( \modularity \).}.

Conversely, if the red vertices are all assigned to modules, maximizing \( \modularity \) consists of assigning the blue vertices to modules. Analogously to the previous case, we define \( \redmodwtmat = \matrixinnerproduct{\modsubmat}{\redindmat} \) and manipulate \eqn{eq:defsubmatmodularity} into the form
\begin{equation}
	\modularity = \modularitybipartnorm \bluesum{j} \largeparens{\modsum{k} \blueindelem{j}{k} \redmodwtelem{j}{k}}
	\mathperiod
	\label{eq:fixedredmodularityrowsums}
\end{equation}
As with the red vertices, we maximize \( \modularity \) by assigning the \( j \)th blue vertex to the module \( k \) such that \( \redmodwtelem{j}{k} \) is the maximum of the \( j \)th row of \( \redmodwtmat \).

Taken together, these two maximization procedures define an algorithm that we call \brim (bipartite, recursively induced modules). The \brim algorithm is an iterative algorithm for maximizing \( \modularity \), with the sets of red and blue vertices each recursively drawing the other into modular structures. For each iteration, \( \modularity \) is guaranteed never to decrease, as it is always possible at least to maintain the previous vertex partitioning and keep the modularity the same. Therefore, the \brim algorithm will always find a partition at a maximum of \( \modularity \). In general, the identified partition will correspond to a local maximum in \( \modularity \), not the global maximum. 

Note that the \brim algorithm can work with the entire \( \modsubmat \) matrix, or a rank-restricted approximation calculated by omitting the smallest singular values. By using the full \( \modsubmat \) matrix, we automatically include all positive contributions to the modularity. As well, the algorithm can work with any assumed number of modules; however, no constraint exists to ensure that each module is occupied. 

To test the efficacy of the \brim algorithm, we apply it to a simple model network. The model consists of \(\nummodelnetmodules\) modules, each containing \( \nummodelnetredpermod \) red and \( \nummodelnetbluepermod \) blue vertices. An edge exists between a red vertex and a blue vertex with probability \( \problinkintra \) if they are in the same module and with probability \( \problinkinter \) if they are in different modules. No edges exist between vertices with the same color. 

The qualitative behavior of the model depends on \( \problinkintra \) and \( \problinkinter \). When \( \problinkintra > \problinkinter \), there is a greater probability of vertices within a module being linked than vertices in different modules, matching our intuitive notion of modularity.  With \( \problinkintra \) sufficiently close to one and \( \problinkinter \) small, the actual modular structure of a particular realization of the model should correspond to the assumed modular structure. As \( \problinkinter \approaches \problinkintra \), the network becomes more uniform, with the assumed modular structure ultimately vanishing and all vertices belonging to a single module.\footnote{More customary  (see, \eg, Ref.~\citep{DanDiaDucAre:2005}) is to fix the expected degree of the network vertices and vary the expected number of edges linking vertices in different modules, with \(\problinkintra\) and \(\problinkinter\) calculated from the expectation values. However, for the bipartite network model under consideration, the base case, with edges only existing between vertices in the same module, will often be excluded using this approach.} Lower values of \( \problinkintra \) introduce additional substructure into the modules; the general behavior as \( \problinkinter \) varies  should be similar to the previous case, but with an overall reduced correspondence between the assumed modules and the actual modules in networks instantiated from the model. 

Following \citet{DanDiaDucAre:2005}, we make precise the above qualitative description in terms of the normalized mutual information \(\normalizedmutinfosymbol\). Consider two schemes \(\schemex\) and \(\schemey\) for dividing the \(\numvertices\) vertices into community groups, represented by two index matrices \(\indmatx\) and \(\indmaty\) \footnote{Analogous measures can be defined in a straightforward fashion using the portions of the index matrices that correspond to just the red or blue vertices.}.  The two index matrices are used to calculate the so-called confusion matrix \(\confusionmat\), which takes the simple form
\begin{equation}
	\confusionmat = \transpose{\indmatx}\indmaty
	\label{eq:defconfusionmatrix}
	\mathperiod
\end{equation}
The probability \(\probability{X=x, Y=y}\) that a vertex is assigned to community \(x\) in scheme \(\schemex\) and to community \(y\) in scheme \(\schemey\) is proportional to the corresponding element \(\confusionelem{x}{y}\) of the confusion matrix, so that
\begin{equation}
	\probability{X=x, Y=y} = \frac{1}{\numvertices} \confusionelem{x}{y}
	\label{eq:relateconfusiontoprobability}
	\mathperiod
\end{equation}
Using the probability as defined in \eqn{eq:relateconfusiontoprobability}, we can calculate the normalized mutual information as
\begin{equation}
	\normalizedmutinfo{X}{Y} = \frac{2\mutinfo{X}{Y}}{\entropy{X} + \entropy{Y}}
	\label{eq:defnormmutinfo}
	\mathperiod
\end{equation}
\Eqn{eq:defnormmutinfo} is expressed in terms of the usual mutual information \(\mutinfo{X}{Y}\) and entropies \(\entropy{X}\) and  \(\entropy{X}\) \citep{CovTho:1991}, defined as
\begin{eqnarray} 
	\mutinfo{X}{Y} & = & \relentsum{\probability{X, Y}}{\probability{X}\probability{Y}}{x,y} \label{eq:defmutinfo}\\
	\entropy{X} & = & \entropysum{\probability{X}}{x} \label{eq:defentropyx}\\
	\entropy{Y} & = & \entropysum{\probability{Y}}{y} \label{eq:defentropyy}
	\mathperiod
\end{eqnarray} 
In \eqns{eq:defnormmutinfo} \througheqn{eq:defentropyy}, we have made use of the common shorthand abbreviations \(\probability{X=x, Y=y} = \probability{X, Y}\), \(\probability{X=x} = \probability{X}\), and \(\probability{Y=y} = \probability{Y}\). The base of the logarithms in \eqns{eq:defmutinfo} \througheqn{eq:defentropyy} is arbitrary, as the computed measures only appear in the ratio in \eqn{eq:defnormmutinfo}. 

The normalized mutual information is a measure of the amount of information common to the two partitioning schemes. By taking one of the partitions to be the assumed modular structure of the network and one to be the structure found using the \brim algorithm, we can thus explore the efficacy of the algorithm. When the found modules match the real ones, we have \(\normalizedmutinfosymbol=1\), and when the found modules are independent of the real ones, we have \(\normalizedmutinfosymbol=0\).

We now set \(\nummodelnetmodules = 5\), \( \nummodelnetredpermod  = 12\), and \( \nummodelnetbluepermod = 8 \), giving \( \numvertices = 100 \) vertices in the network. With various choices of  \( \problinkintra \) and \( \problinkinter \), we repeatedly instantiate the model network and determine the assignment of vertices to modules using the \brim algorithm. The algorithm is initialized by assigning each of the blue vertices to a unique module. For each sample, we calculate \(\normalizedmutinfosymbol\).

\begin{figure}[htbp]
	\includegraphics[width=\columnwidth]{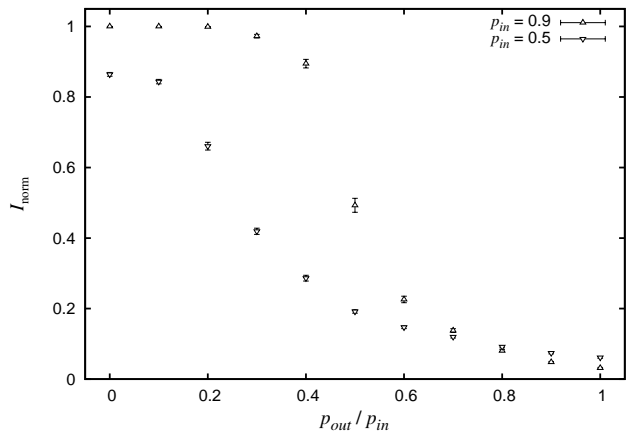}
	\caption{Agreement between model network modules and modules found using the \brim algorithm. Each point shows the mean normalized mutual information between the model network community groups and those identified using the algorithm, averaged over 100 realizations of the model network. Error bars show the standard error of the mean.}
	\label{fig:validatebrim}
\end{figure}

In \fig{fig:validatebrim}, we show results of applying the \brim algorithm to the model network. The points show the mean value of  \(\normalizedmutinfosymbol\), averaged over 100 instantiations of the network. The error bars show the standard error of the mean. The general behavior is as anticipated, lending confidence to the algorithm definition. 

% subsection recursive_identification_of_bipartite_modules (end)

\subsection{Determining the Number of Modules} % (fold)
\label{sec:num_modules}

The \brim algorithm is silent on the issue of how many modules should be used. As noted in \sxn{sec:module_identification_in_bipartite_networks}, the number of modules \( \nummodules \) is at most one more than the rank of \( \modsubmat \), which is a relatively weak constraint. One approach is thus to assign each vertex of the smaller of the red and blue vertex sets to unique modules, and allow the vertices to be grouped into an appropriate number of modules. For the \brim algorithm, said approach is resource intensive, requiring the calculation of modularity contributions for what may be a grossly overestimated number of modules. Worse still, when the number of vertices is much greater than the number of modules, the \brim algorithm may terminate at low-quality local maxima far from the true number of modules in the network (see \sxn{sec:scotland} for an example of this).

Clearly, automatically selecting the correct number of allowed modules in such a case would be preferable. The allowed number of modules \( \nummodules \) thus becomes an adaptable parameter for which a value is to be found that optimizes the modularity. This presents some difficulties in that there is no obvious relationship between the allowed number of modules and the modularity found by the \brim algorithm. However, by assuming that the modularity depends on the allowed number of modules in a reasonably smooth fashion, we can use a simple bisection approach to identify an appropriate value for the number of allowed modules.

The search begins by requiring all vertices to belong to the same module, \( \nummodules = 1 \), giving \( \modularity = 0 \). We double the allowed number of modules \( \nummodules \).  Half of the vertices are randomly reassigned to the newly defined modules, and a new, locally optimal solution is found using the \brim algorithm. This process continues, with \( \nummodules \) being repeatedly doubled so long as \( \modularity \) continues to increase. Each step in the \( \nummodules \)-search builds on the previous solution by partially reusing the assignment of vertices to modules.

Once \( \modularity \) drops as \( \nummodules \) increases, we have crossed a maximum in the modularity landscape. We therefore switch from extrapolating to larger numbers of modules to interpolating within the interval that includes the maximum. The interpolation is done using a simple bisection search in the allowed number of modules, trying new values for \( \nummodules \) so as to continuously reduce the interval wherein the putative maximum in \( \modularity \) lies. As with the initial extrapolation stage of the search, vertices are assigned from earlier solutions to the newly allowed modules for each value of \( \nummodules \), and a new, locally optimal solution found. 

The search for \( \nummodules \) terminates once the interval becomes sufficiently small. In this work, we take the interval to be 2, \ie, the \( \modularity \) maximum at \( \nummodules = \nummodulesmax \) is bracketed by inferior solutions at \( \nummodules = \nummodulesmax - 1 \) and \( \nummodules = \nummodulesmax + 1 \).  This adaptive \brim algorithm enables us to identify the appropriate number of modules \( \nummodulesmax \) in a number of steps that scales logarithmically with the number of vertices in the network.

% subsection num_modules (end)

% section module_identification (end)

\section{Results}\label{sec:results} % (fold)

In this section, we apply the \brim algorithm to a network showing the interactions of women in the American Deep South at various social events \citep{DavGarGar:1941} and to a network showing corporate interlocks in Scottish firms \citep{ScoHug:1980}. Both networks are conveniently available on the World Wide Web in Pajek format \citep{BatMrv:2006}.

\subsection{Southern Women Event Participation}\label{sec:southern_women} % (fold)

As an initial example, we consider the Southern women data set, collected by \citet{DavGarGar:1941} in and around Natchez, Mississippi during the 1930s as part of an extensive study of class and race in the Deep South. This data set and networks derived from it have been much studied. Indeed, \citet{Fre:2003} has described it as ``\ldots a touchstone for comparing analytic methods in social network analysis.''

The Southern women data set describes the participation of 18 women in 14 social events. The women and social events constitute a bipartite network; an edge exists between a woman and a social event if the woman was in attendance at the event. The network is connected.

We identified network modular structure using the \brim algorithm. The initial state is in general important. The dependence on the initial state is most visible in the quality of the stable solution, i.e., the algorithm can get ``stuck'' at a poor quality local maximum. We initialized the assignment of events to modules in  \( \blueindmat \) using several strategies: (1) assigning all events to a single module, (2) assigning each event to its own module, and (3) randomly assigning events to modules. 

For this network, all three strategies identify modular structures. The first strategy produces a good quality solution (4 modules, \( \modularity = 0.34554 \)). The second strategy also produces a solution that captures a great deal of the modular structure, but is somewhat coarser than the first (2 modules, \( \modularity = 0.32117 \)). The third strategy, random initial assignment, sheds light on the quality of the first two. Because the network is small, a large number of trials can be run without difficulty; we ran 500,000 trials. The greatest modularity found equalled that found with all events initially in unique modules, \( \modularity = 0.34554 \), indicating that this best solution found is quite good.

\begin{figure}[htbp]
	\includegraphics[width=\columnwidth]{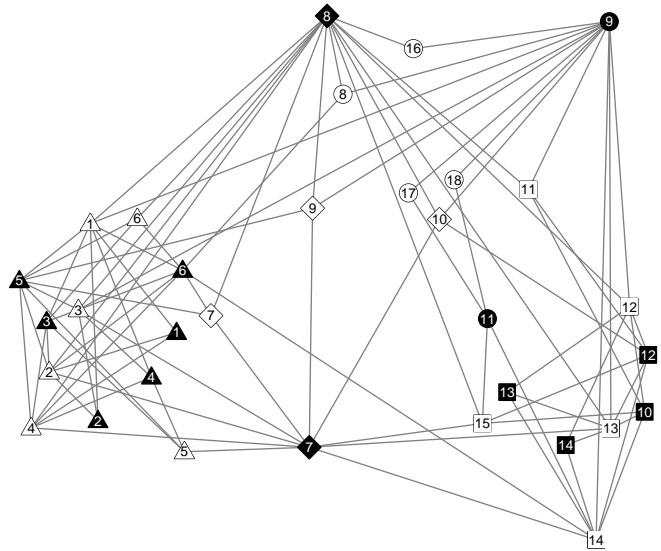}
	\caption{Modules in the Southern women network. The women are represented as open symbols with black labels and the events as filled symbols with white labels. The modules are indicated by the shape of the symbols. Vertices are positioned with coordinates based on the elements of the singular vectors corresponding to the two largest singular values of \( \modsubmat \); some vertices are repositioned slightly to eliminate overlaps. The vertex partition pictured has the highest modularity we have found for the Southern women network, \( \modularity=0.34554 \)}
	\label{fig:southernwomengraph}
\end{figure}

In \fig{fig:southernwomengraph}, we show the best assignment of vertices to modules determined using the \brim algorithm with all events initially in different modules. The shapes of the vertices show which ones belong to the same modules, with four modules in all. Open symbols with black labels portray vertices corresponding to the women, and filled symbols with white labels portray vertices corresponding to the events. The positions of the vertices are based on the singular vectors corresponding to the two largest singular values of \( \modsubmat \), with the right singular vectors giving the coordinates for the events and the left singular vectors giving the coordinates for the women. Several vertices have been shifted slightly to prevent overlapping vertex symbols while preserving the overall character of the network.

The community groups found using the \brim algorithm are comparable to those found in previous investigations of the Southern women data set (Ref.~\cite{Fre:2003} provides a useful survey). Most such studies have focused on the women, leaving the groupings for the events unspecified; we can use the groupings of the women to assign the events to the best modules, as described in \sxn{sec:recursive_identification_of_bipartite_modules}, and calculate modularity values for purposes of comparison. The community groups can be further compared using the normalized mutual information between the various groupings of the women and the best grouping found using the \brim algorithm. Values of \(\modularity\) and \(\normalizedmutinfosymbol\) are summarized in \tbl{tab:southernwomen} and discussed in depth below.

\begin{table}
\caption{Comparison of modules in the Southern women network. Where necessary, the modularity values \(\modularity\) are calculated from an optimistic assignment of the events to the best possible modules from a given assignment of the women to modules. Values of the normalized mutual information \(\normalizedmutinfosymbol\) are calculated between the given divisions of the women and the best division found using the \brim algorithm.  }\label{tab:southernwomen}
\begin{ruledtabular}
\begin{tabular}{ldd}
	Modules & \multicolumn{1}{c}{\modularity} & \multicolumn{1}{c}{\normalizedmutinfosymbol} \\ \hline
	\brim & 0.34554 &  1 \\
	Spectral & 0.32117 & 0.56897 \\
	Davis 1 & 0.31057 & 0.44657 \\
	Davis 2 & 0.31839 & 0.45126 \\
	Doreian & 0.29390 & 0.60766 \\
	Unipartite & 0.21866 & 0.28019 \\
\end{tabular}
\end{ruledtabular}
\end{table}

In the original investigation, \citet{DavGarGar:1941} used general ethnographic knowledge of the community to assign the women to two groups. The groups consisted of women 1--9 and of women 9--18; woman 9 is a secondary member of both groups. To be consistent with the definitions in \sxn{sec:bipartite_modularity}, we must assign this individual to a specific group. The \(\modularity\) and \(\normalizedmutinfosymbol\) values are seen from \tbl{tab:southernwomen} to be similar for both assignments, with the case where woman 9 is grouped with women 10--18 labeled as ``Davis 1'' and the case where woman 9 is grouped with women 1--8 labeled as ``Davis 2.'' The latter division is the same as what \citet{Fre:2003} identified as the consensus from 21 different studies of the Southern women data set. The \(\modularity\) and \(\normalizedmutinfosymbol\) values are reasonably similar to values found for two modules using either the \brim algorithm or spectral bipartitioning as discussed in \sxn{sec:module_identification_in_bipartite_networks}, which groups the women into sets \{1--7, 9\} and \{8, 10--18\} (identified in \tbl{tab:southernwomen} with the label ``Spectral''). 

\citet{DorBatFer:2004} considered the modular nature of both parts of the network, suggesting several divisions of the women and events. The division with the greatest modularity (given in their Table 4) is characterized in \tbl{tab:southernwomen} with the label ``Doreian.''  Taking just their partitioning of the events into three groups (events 1--5, 6--9, and 10--14) and replacing their partitioning of the women using the approach from \sxn{sec:recursive_identification_of_bipartite_modules}, the modularity can be increased from 0.29390 to 0.32950. This is similar to the best assignment of vertices to modules we described above, with modularity of 0.34554, wherein the additional structure produces a modest, but real, improvement in the modularity. 

It is also of interest to compare the community groups obtained for the Southern women network using the bipartite network to those found using an unweighted projection network. Here, we focus on the projection consisting of the eighteen women as vertices, with edges defined by mutual participation in events. The best division we found for the women, discussed above and shown in \fig{fig:southernwomengraph}, actually has a negative value for the standard unipartite modularity; it is thus better to use only a single module containing all eighteen women than the best module found for the bipartite network. Since the modules we identified from the bipartite network using the \brim algorithm are similar to those found in numerous other studies, this highlights the difficulties that can arise using a unipartite projection.

Conversely, we can determine the bipartite modularity for community groups found using the unipartite projection. We first use the Newman vector to partition the women into two groups as described in \sxn{sec:spectral_methods_for_module_identification}, with women 2 and 4--7 in one group and all others in a second group. Next, we determine the best assignment of events to modules using the approach from \sxn{sec:recursive_identification_of_bipartite_modules}. Together, this gives the values shown in \tbl{tab:southernwomen} for the label ``Unipartite,'' which reflect that some of the modular structure of the network has been captured but is generally inferior to the solutions found from the bipartite network. Further, the solution from the unweighted projection does not correspond to a maximum in the bipartite modularity; using the solution as the initial state for the \brim algorithm, a solution is obtained with two modules identical to those found using spectral bipartitioning as described in \sxn{sec:module_identification_in_bipartite_networks}.

% subsubsection southern_women (end)

\subsection{Scotland Corporate Interlock}\label{sec:scotland} % (fold)

As a second example, we consider a data set on corporate interlocks in Scotland in the early twentieth century \citep{ScoHug:1980}. The data set characterizes 108 Scottish firms during 1904-5, detailing the corporate sector, capital, and board of directors for each firm. The data set includes only those board members who held multiple directorships, totaling 136 individuals. 

Here, we focus on the bipartite network of firms and directors, with edges existing between each firm and its board members. Unlike the Southern women network, the Scotland corporate interlock network is not connected. In the following, we consider only the largest component of the graph, containing 131 directors and 86 firms---and thus, as many as 86 modules.

\begin{figure}[htbp]
	\includegraphics[width=\columnwidth]{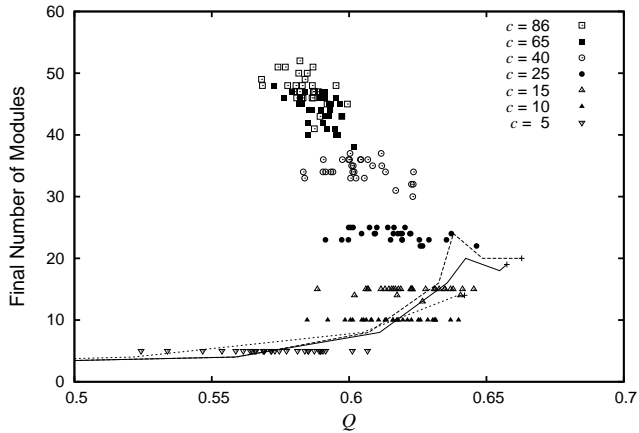}
	\caption{Quality of solutions found in the Scotland corporate interlock network. The modularity \( \modularity \) depends on the allowed number of modules \( \nummodules \). The points correspond to solutions found using the \brim algorithm starting from a random initial assignment of vertices to modules. The values on the ordinate indicate the number of modules occupied by at least one vertex in the solution state found by the \brim algorithm. All points are slightly dithered to better show regions with many similar or identical solutions.  The lines show the course of an adaptive search for the correct number of modules to maximize the modularity, terminating at states with the modularity and number of modules shown by the crosses.}
	\label{fig:brim_results}
\end{figure}

As with the Southern women network, assigning all directors to unique modules or to the same module results in a solution that captures some of the modular character of the network, with \( \modularity = 0.56634 \) and \( \modularity = 0.39873 \), respectively. However, in contrast to the Southern women network, these are rather poor solutions to what can be found starting from a random assignment of directors to modules (see \fig{fig:brim_results}).

Further, the best solutions are found by restricting the allowed number of modules \( \nummodules \) to less than the maximum. In principle, allowing the number of modules to take on any size leaves the \brim algorithm to search the largest possible space, potentially finding the largest possible modularity value. In practice, the results are inferior to those obtained from a more restricted search.  In \fig{fig:brim_results}, we show the results, in terms of the actual numbers of modules occupied and modularity values, for \brim searches with the allowed number of modules restricted. This trades off the possibility of higher modularity values in the excluded region for improved searching in the remaining region. The trade-off is clearly a good one, as the best solutions are found with fewer than thirty modules. 

In \fig{fig:brim_results}, we also show three runs of the adaptive \brim algorithm described in \sxn{sec:num_modules}. The lines show the progress of the number of modules and modularity value during the search. The number of modules \( \nummodules \) allowed for the \brim search is typically close (within 10\%) to the number of modules actually found, suggesting that the adaptive approach eliminates a wasteful search through vertex assignments with too many modules. The three traces all show typical behavior and lead to good solutions; two of the adaptive runs lead to better solutions, in terms of modularity, than any of the much larger number of trials using \brim with a fixed \( \nummodules \).

Based on the solutions shown in \fig{fig:brim_results}, the main component of the Scotland corporate interlock network has roughly twenty community groups, considerably fewer than the 131 directors or 86 firms. This analysis could serve as a starting point for an investigation of the community structures of the firms or directors. A more comprehensive analysis would take into account the available information on the corporate sectors and capital of the firms.

% subsubsection scotland (end)

% section results (end)

\section{Conclusions}\label{sec:conclusions} % (fold)

We have defined and explored a modularity appropriate for bipartite networks. The presented results extend and specialize the matrix-based approach recently reported by \citet{New:2006a} for unipartite networks. The bipartite structure of the network is reflected mathematically in the importance of an asymmetric submatrix \( \modsubmat \) of the full bipartite modularity matrix \( \modmat \), with a corresponding emphasis on the singular value decomposition of \( \modsubmat \) instead of the spectral decomposition of \( \modmat \). We made use of the properties of \( \modsubmat \) to define an algorithm, \brim, for use in identifying network modules. By applying the algorithm to real-world networks, we demonstrated its effectiveness and identified some of its limitations. 

The usual unipartite modularity has a limited resolution that depends on the number of edges in the network \citep{ForBar:2007}. The main consequence of the resolution limit is that the modules in large networks may have hidden substructures that require deeper investigations to reveal. Although we have not shown it, we expect that the bipartite modularity introduced in this work has a similar resolution limit, with similar consequences.

One of the key themes in this paper has been that the bipartite structure of the network can be beneficially incorporated into its mathematical description and its computational treatment. This theme was realized in the \brim algorithm, where the assignment of vertices to modules in one part of the network, when held fixed, provides a stable modularity landscape in which it is straightforward to partition the vertices of the other part into modules. We expect that the characteristics of other specialized classes of networks could be taken advantage of in an analogous fashion to define appropriate null model networks, modularity measures, and community detection algorithms. 

The eigenvalues of the graph Laplacian are closely related to many important properties and invariants of the graph \citep{Chu:1997}. In contrast, relatively little is known about the spectra of  modularity matrices, be they for unipartite or bipartite networks. We are optimistic that the eigenvalues of the modularity matrix usefully relate to important and interesting network properties.

% section conclusions (end)

\begin{acknowledgments}

The author thanks Ludwig Streit, Philippe Blanchard, and Thomas Roediger-Schluga for useful comments and suggestions. This work has been supported in part by the European FP6-NEST-Adventure Programme, contract number 028875.

\end{acknowledgments}

%: Bibliography

\bibliographystyle{apsrev}

\end{document}